\documentclass[amsmath,twocolumn,pra,longbibliography]{revtex4-2} 
\usepackage{amsthm,amssymb,amsfonts,graphicx,verbatim, xcolor,bm} 
\usepackage{hyperref} 
\usepackage[utf8]{inputenc} 
\usepackage{siunitx} 

\usepackage{tikz} 
\usetikzlibrary{arrows,decorations.pathmorphing,backgrounds,positioning,fit,matrix}

\let\oldmarginpar\marginpar
\renewcommand\marginpar[1]{\-\oldmarginpar[\raggedleft\footnotesize #1]%
{\raggedright\footnotesize #1}}
\renewcommand{\eqref}[1]{(\ref{#1})}

\renewcommand{\vec}[1]{{\bf #1}}

\renewcommand{\Re}{{\rm Re}\,}

\newcommand{\ket}[1]{|#1\rangle}

\newcommand{\braOket}[3]{\langle #1|#2|#3\rangle}

\begin{document}

\title{
New mechanism and exact theory of superconductivity from strong repulsive interaction
}
\author{Valentin Cr\'epel, Liang Fu}
\affiliation{Massachusetts Institute of Technology, 77 Massachusetts Avenue, Cambridge, MA, USA}

\begin{abstract}
{\it Abstract:} 
We introduce a new and general mechanism for superconductivity in Fermi systems with strong repulsive interaction. Because kinetic terms are small compared to the bare repulsion, the dynamic of charge carriers is constrained by the the presence of other nearby carriers. 
By treating kinetic terms as a perturbation around the atomic limit, we show that pairing can be induced by correlated multi-particle tunneling processes that favor two itinerant carriers to be close together. Our analytically-controlled theory provides a quantitative formula relating $T_c$ to microscopic parameters, with maximum $T_c$ reaching about 10$\%$ of the Fermi temperature. 
Our work demonstrates a powerful method for studying strong coupling superconductivity with unconventional pairing symmetry. It also offers a realistic new route to realizing finite angular momentum superfluidity of spin-polarized fermions in optical lattice.

{\it One-sentence summary:} We present a new theory of superconductivity with pairing induced by correlated tunneling processes involving three particles.
\end{abstract}

\maketitle

\section{Introduction}

Superconductivity in conventional metals results from an effective attraction between electrons mediated by the exchange of phonons~\cite{bardeen1957theory,frohlich1950theory}.
While this attraction is much weaker than the bare Coulomb repulsion~\cite{bardeen1955electron}, the latter is drastically renormalized downward by retardation effects~\cite{morel1962calculation}. 
Thanks to the vast difference between Fermi and Debye energy, the phonon-mediated attraction can overscreen the Coulomb repulsion to enable electron pairing and superconductivity~\cite{bogoljubov1958new}. On the other hand, this crucial retardation condition fails in systems with narrow bands or low carrier density. Yet superconductivity has been found in a growing number of materials in such strong-coupling regime. Two famous examples are (1) strontium titanate, the most dilute bulk superconductor with Fermi energy as small as $1$meV~\cite{PhysRevX.3.021002};  (2) magic-angle graphene with a record-low density $n_{\rm 2D} \sim 10^{11} \SI{}{\per\centi\meter\squared}$ and a very small bandwidth $\sim 10$meV~\cite{cao2018unconventional,lu2019superconductors,park2020tunable,hao2020electric}. Remarkably, the ratio of superconducting transition temperature $T_c$ and Fermi temperature $E_F/k_B$ far exceeds typical values, reaching as high as 0.01 in strontium titanate~\cite{PhysRevX.3.021002} and 0.1 in magic-angle graphene  ~\cite{lu2019superconductors}. 
Finding electronic mechanisms for strong-coupling superconductivity in narrow band systems has long been a subject of great interest and challenge~\cite{norman2011challenge,phillips1998superconductivity,ruhman2016superconductivity,qin2020absence,raghu2010superconductivity,maiti2013superconductivity,khaliullin2008origin,nandkishore2014superconductivity}.

In this work, we introduce a new mechanism for superconductivity stemming from the strong repulsive interactions.
Because kinetic terms are small compared to the bare repulsion, the dynamic of charge carriers is constrained by the the presence of other nearby carriers. 
In this regime, pairing can be induced by correlated multi-particle tunneling processes that favor two itinerant carriers to be close together. 
To illustrate this physical phenomenon, we introduce a simple two-band model of interacting spin-polarized fermions on a two-dimensional lattice.
In our model, an insulating state occurs at the filling of $n=1$ fermion per unit cell, and superconductivity emerges  upon particle or hole doping. 
Based on a perturbative expansion around the atomic limit, we rigorously show that, despite the strong bare repulsion, a  non-retarded short-range pairing interaction between doped fermions arises. It is generated by coupling to high-energy composite excitations, which mediate correlated-tunneling terms, effectively keeping pair of carriers close to one another.
The resulting superconductor is unconventional by all standards.  
It has $f$-wave pairing symmetry and changes  from having a full gap to point nodes above a critical doping. $T_c$ is controlled by the bare interaction strength and the band gap, reaching as large as $T_c \sim 0.1 E_F/k_B$ when they are of comparable magnitude.
Our  theory is analytically controlled by a small coupling constant that emerges in the narrow band limit.
Our work demonstrates a reliable and robust mechanism for unconventional superconductivity from repulsion in narrow band systems.

Our model is broadly inspired by a recent work by Slagle and one of us~\cite{slagle2020charge}, who proposed a mechanism for pairing from purely {\it classical} electrostatic repulsion in a doped charge transfer insulator. 
The essential ingredient there is a charge-$2e$ excitation dubbed ``trimer'', which is a composite object  
consisting of two doped electrons  tightly bound to a dipole. 
Interestingly, for certain extended Coulomb repulsion, a trimer is energetically more favored than two separate electrons.
Under appropriate conditions, the presence of preformed trimers can lead to Wigner crystal or superconducting ground states at small doping.

While we also start with the problem of doping an insulator, our work differs fundamentally from Ref.~\cite{slagle2020charge}. 
We find superconductivity without invoking trimers or any preformed pairs at low energy.
Instead, pairing arises from correlated {\it quantum} hopping of doped fermions induced by {\it virtual} composite excitations at high energy. 
We demonstrate this novel mechanism by introducing and solving the simplest model 
of spin-polarized fermions interacting on a bipartite lattice. 
Last but not the least, our solution reveals distinct superconducting states at different ranges of doping and provides a quantitative formula for $T_c$ in terms of microscopic parameters.

\section*{Results}

We consider spin-polarized fermions on a bipartite lattice with repulsive interactions, described by the Hamiltonian
\begin{eqnarray} \label{eq:OriginalModel} 
\mathcal{H} &=& \mathcal{H}_0 + \mathcal{H}_t, \nonumber \\ 
\mathcal{H}_0  &=& V \sum_{\langle r, r' \rangle} n_{r} n_{r'}  + \Delta \sum_{r\in B} n_r, \nonumber \\
\mathcal{H}_t &=& - t \sum_{\langle r, r' \rangle} (c_{r}^\dagger c_{r'}  + {hc}).
\end{eqnarray}
where $\mathcal{H}_0$ contains the nearest-neighbor interaction -- the dominant interaction for spin-polarized fermions on a lattice -- and the sublattice potential difference between the two inequivalent $A$ and $B$ sites, while $\mathcal{H}_t$ describes tunneling between adjacent sites. 
Despite the simplicity of our model, in this work we unveil its remarkably rich phase diagram as a function of filling and interaction strength.
We shall  derive the low-energy properties of the system with a fully-controlled perturbative expansion in the narrow-band limit $t \ll \Delta$, which we further complement with field-theoretic analysis and extensive exact diagonalization (ED) studies. For concreteness, we thereafter focus on the honeycomb lattice.

\begin{figure}
\centering
\includegraphics[width=\linewidth]{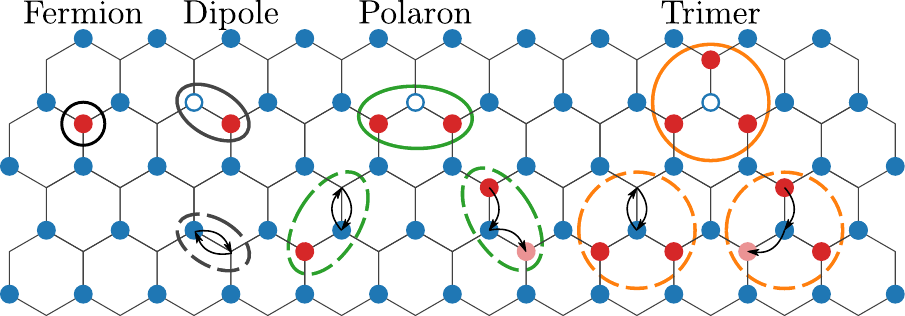}
\caption{\textbf{Model and virtual processes:}
Low energy fermion added above the $n=1$ insulating background live on the $B$ lattice. Excitations above this $f$-band are dipoles, polarons and trimers (solid circles), whose virtual occupation leads to the effective model Eq.~\eqref{eq:BelectronEffModel} (dashed circles).
}
\label{fig:HoneycombLattice}
\end{figure}

In the strong coupling limit, the ground state of $\mathcal{H}$ at $n=1$ is an insulator with all $A$ sites occupied and all $B$ sites empty, as shown in Fig.~\ref{fig:HoneycombLattice}. 
Its insulating property is ensured by the large gap $E_{D} = 2V+\Delta$, which corresponds to the energy necssecary to transfer an electron from an $A$ site to $B$, or equivalently, creating a dipole. The inclusion of tunneling, small compared to $E_{D}$, slightly decreases the charge transfer gap without any significant change to the insulating ground state.

Since $\mathcal{H}$  is invariant under particle-hole transformation $c_{A} \rightarrow c_A^\dagger, c_B \rightarrow - c_B^\dagger$ combined with spatial inversion that interchanges the two sublattices, it suffices to consider $n>1$ filling below. 
At finite doping $n=1+\delta$ ($\delta>0$), low energy configurations of the system remain with all $A$ sites occupied in order to avoid the large charge transfer gap. Due to Pauli exclusion principle, the $\delta$ additional fermions must live on the $B$ lattice, and in the limit $t=0$, form a highly degenerate manifold with an energy $E_f = \Delta + 3V$ per doped charge that we refer to as $f$-band.

Besides these fermions on $B$ sites, there exist various types of composite excitations at higher energy, which involve holes on $A$ sites, as depicted in Fig.~\ref{fig:HoneycombLattice}. For example, a $B$-fermion can bind with a neighboring dipole to form a charge-$e$ Fermi polaron, which has energy $E_P= E_f + V+ \Delta$. More interesting is the charge-$2e$ trimer, which consists of three neighboring $B$-fermions surrounding a hole on the center $A$ site. It can also be viewed as two neighboring $B$-fermions tightly bound to a dipole. A trimer costs energy $E_T= 2E_f + \Delta$, which is greater than the energy of two separate $B$-fermions by $\Delta$. These composite excitations -- dipoles, polarons and trimers -- are hereafter collectively referred to as charge-transfer complex.

In the presence of small quantum tunneling $t \ll \Delta$, doped carriers in the $f$-band constitute the only low-energy excitations in our system. They virtually couple to charge-transfer complex at high energy. This coupling results in a narrow dispersive $f$-band of doped carriers and induces short-range interactions between them.
Remarkably, we shall show  that the induced interaction leads to pairing within the $f$-band. 
To that purpose, we analytically carry out a Schrieffer-Wolff transformation $\mathcal{H}' = e^{iS} \mathcal{H} e^{-iS}$ to decouple the $f$-band from high energy degrees of freedom~\cite{schrieffer1966relation,MacDonald_tOverVExpansionHubbard}. 
As detailed in the supplementary materials~\cite{SuppMat}, this procedure accounts for all possible virtual processes (see Fig.~\ref{fig:HoneycombLattice}), and leads to the following effective Hamiltonian for doped fermions,  which is {\it exact} to second order in $t/\Delta$ and at any $f$-band filling:
\begin{equation} \label{eq:BelectronEffModel} \begin{split}
\mathcal{H}' = &  \sum_{\langle i, j \rangle} t_f ( f_i^\dagger f_j  + hc ) + V_f n_i n_j \\
& + \sum_{ (ijk) \in \triangle } \lambda (f_i^\dagger n_j f_k + P_{ijk}) + U_3 n_i n_j n_k   .
\end{split} \end{equation}
The $f_i$ fermionic operators denote the doped fermions on the triangular $B$-lattice, and their vacuum is the $n=1$ insulating state described above. 
The sums labeled by $\langle i, j \rangle$ and $(i,j,k) \in \triangle$ respectively run over all bonds and all {\it upper} triangles of the $B$-lattice, while $P_{ijk}$ stands for the inclusion of $f_i^\dagger n_j f_k$ with all possible permutations of the indices $i$, $j$ and $k$.

The effective Hamiltonian $\mathcal{H}'$ for doped fermions consists of  single-particle tunneling, correlated (density-dependent) tunneling, two-body and three-body density interactions.
Their origins can be understood as follows.
The tunneling from $k$ to $i$ in the upper triangle $(ijk)$ arises from two consecutive hopping processes.
The virtual intermediate state involved is either a polaron or a trimer, depending on the occupation of site $j$ (see Fig.~\ref{fig:HoneycombLattice}). The  resulting tunneling amplitude is thus $t^2 [ (1-n_j) / (E_P-E_f) + n_j/(E_T-2E_f)]$, from which the expression of $t_f$ and $\lambda$ are derived:
\begin{subequations} \label{eq:CoefficientSchriefferModel}
\begin{align}
t_f & = \frac{t^2}{\Delta + V}  \, , \\ 
\lambda & = \frac{t^2}{\Delta} - \frac{t^2}{\Delta + V} \, .
\end{align}
The interaction coefficients $V_f$ and $U_3$ come from processes where an $A$ fermion hops back and forth between neighboring sites (see Fig.~\ref{fig:HoneycombLattice}). 
For example, $V_f$ measures the difference of energy between two neighboring doped charges on $B$ sites and two well-separated ones. The former configuration can couple to a trimer state whereas the latter cannot, thus leading to an attraction $-t^2/\Delta$. 
Accounting for all processes, we find
\begin{align}
V_f & = -\frac{t^2}{\Delta} + \frac{4t^2}{\Delta + V} - \frac{3t^2}{\Delta + 2V} \, , \\
U_3 & = \frac{3t^2}{\Delta} - \frac{6t^2}{\Delta + V} + \frac{3t^2}{\Delta + 2V} \, .
\end{align} 
\end{subequations}
In Eq.~(\ref{eq:CoefficientSchriefferModel}a-d), the denominators $\Delta$, $\Delta+V$ and $\Delta + 2V$ are the energy costs of intermediate states involving  trimer, polaron and dipole respectively. When longer-range interactions are included, these energy denominators change accordingly. Importantly, higher-order corrections to the effective Hamiltonian $\mathcal{H}'$ are small provided that the narrow band condition $t\ll \Delta$ is satisfied, regardless of interaction strength $V$.

At small $V\ll \Delta$, the effective interactions in the $f$-band are found to be $V^0_f =  2(t/\Delta)^2 V, \lambda^0=(t/\Delta)^2 V$ and $U^0_3=0$ to first order in $V/\Delta$. These values simply correspond to the projection of the bare repulsion $V$ into the $f$-band, whose wavefunctions have small amplitudes $\sim t/\Delta$ on $A$ sites.    
As $V$ increases, interband mixing quickly becomes important and our exact results (\ref{eq:CoefficientSchriefferModel}a-d) reveals a dramatic departure of the ``dressed'' interaction from the projected interaction. 
As opposed to the projected interactions, $V_f$ starts to decrease at $V\approx 0.29\Delta$ and changes sign from repulsive to attractive at $V=\Delta$. 
Similarly, $\lambda$ and $U_3$ show manifest deviations from the projected estimates for $V>0.1\Delta$, and saturate at large $V/\Delta$~\cite{SuppMat}. In the rest of this work, we shall mainly consider the case $V<\Delta$, where the induced $V_f, \lambda, U_3$ turn out to be positive and small compared to the single-particle bandwidth $W=9t_f$.

\begin{figure}
\centering
\includegraphics[width=\linewidth]{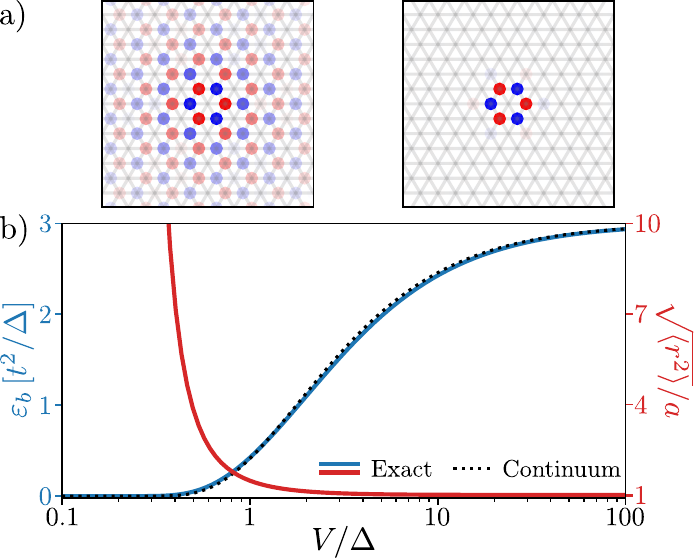}
\caption{\textbf{Doped charges form bound pairs:}
a) The transition from weakly to tightly bound pairs, depicted for $V/\Delta = 0.5$ (left) and $10$ (right), constitutes a probe of the BCS-BEC crossover at low doping. 
b) Pair binding energy $\varepsilon_b$ and bound state size as a function of $V/\Delta$ from the exact solution of the two fermion problem. The prediction of the continuum model (dashed) perfectly match the full-fledged lattice calculation, a stringent test of our derivation. 
}
\label{fig:PairBindingBandwidth}
\end{figure}

To reveal the tendency towards pairing, we study the formation of two-particle bound states, the analog of ``Cooper problem'' in a doped insulator.  Pair formation is evidenced by the positivity of the pair binding energy 
\begin{eqnarray} \label{eq:FirstPairBinding}
\varepsilon_b &=& 2[E(1)-E(0)] - [E(2) - E(0)] \nonumber \\
&=& 2E(1) - E(2) - E(0) ,
\end{eqnarray}
with $E(m)$ the ground state energy of a system with $m$ charges added above $n=1$ filling. As detailed in the supplementary materials~\cite{SuppMat}, we analytically solve the lattice  Hamiltonian Eq.~\eqref{eq:BelectronEffModel} in the case of two doped fermions and obtain $\varepsilon_b$ as a function of $V/\Delta$, shown in Fig.~\ref{fig:PairBindingBandwidth}.
It is found positive in the entire range $V/\Delta$. 
This result shows an effective pairing interaction between low-energy fermions.
It is worth noting that the two-particle bound states cannot be captured by the projected interaction $V_f^0=2\lambda^0$, which proves that pairing is induced by virtual interband excitations.
We further confirm the formation of pairs in the original model Eq.~\eqref{eq:OriginalModel} with ED, see~\cite{SuppMat}. Interestingly, pairs are already present for $V<\Delta$ despite a repulsive induced nearest neighbor interaction ($V_f>0$). This highlights the essential role of correlated hopping $\lambda$ in the effective model for pairing.

The two-particle bound state we found has zero total momentum, is symmetric under three-fold rotation, and changes sign under reflection that flips one of the primitive vectors $\vec{a}_j$, \textit{i.e.} it has $f$-wave pairing symmetry.   
The size of the bound state shrinks with increasing $V/\Delta$, as shown in the inset of Fig.~\ref{fig:PairBindingBandwidth} for $V/\Delta=0.5$ and $100$.  At small $V/\Delta$, the pair wavefunction is highly extended over many lattice sites, while in the opposite limit $V/\Delta \rightarrow \infty$, two nearest-neighbor fermions form the most tightly bound pair ``resonating'' within a single upper triangle (see~\cite{SuppMat}). Our solution of the two-particle problem suggests a crossover between BCS and BEC states at small particle density, tuned by $V/\Delta$.

The presence of a two-particle bound state is a striking feature of our mechanism, distinct from the Kohn-Luttinger mechanism~\cite{Kohn_NewMechanism}, which relies on the presence of a filled Fermi sea to mediate an effective attraction between electrons. Here on the contrary, pairing is already present for two doped fermions, and is induced by virtual excitation of the insulating state.

We now extend our analysis to finite, but small, doping concentrations $\delta$ above $n=1$. 
In this regime, the physics of dilute doped fermions is governed by long-wavelength properties that transcend the details on the lattice scale. This motivates us to derive a low-energy theory by taking the continuum limit of the lattice Hamiltonian. This is achieved by rewriting the lattice Hamiltonian in terms of fermionic fields in momentum space and retaining only modes near the bottom of the $f$-band. Importantly, the band dispersion $\varepsilon(\vec{k}) = 2 t_f \sum_{j=1}^3 \cos ( {\bf k} \cdot {\bf a}_j )$,  with $t_f>0$, has two degenerate minima located at the $\pm \bf K$ points of the  Brillouin zone. Therefore, low-energy degrees of freedom are described by two long-wavelength fermionic fields $\psi_\tau({\bf q}) = f(\tau {\bf K}+ {\bf q})$ with $qa \ll 1$, distinguished by the valley index $\tau = \pm$.
In the supplementary materials~\cite{SuppMat}, we find the following continuum Hamiltonian for $\psi_\tau$:
\begin{equation} \label{eq:ContinuumModel}
\widetilde{\mathcal{H}} = \int {\rm d}x \, \sum_{\tau = \pm} \psi_\tau^\dagger  \left[ \frac{-\nabla^2}{2m} \right] \psi_\tau + g \psi_+^\dagger \psi_+ \psi^\dagger_- \psi_- , 
\end{equation}
where the effective mass and interaction strength are entirely determined from the lattice parameters:  
\begin{eqnarray}
\label{couplingconstant}
m&=& 2/(3t_f a^2), \nonumber \\
g &=&  6 a^2 (V_f - 2 \lambda) <0. 
\end{eqnarray}

The resulting quantum field theory describes a {\it two-flavor} fermion gas in the continuum with \emph{attractive} contact interaction. 
The two flavors correspond to the valley degree of freedom associated with the underlying lattice, from which the field theory is derived. 
This theory is asymptotically exact in the low doping limit where $s$-wave scattering between fermions of opposite valleys is the dominant interaction.

This attractive interaction leads to the formation valley-singlet two-particle bound states, which exactly correspond to the $f$-wave pairs observed on the lattice (see Fig.~\ref{fig:PairBindingBandwidth}). Indeed, the pair amplitude $f_{+K} f_{-K}$ is odd under the reflection that interchanges the two valleys.
To verify the validity of our continuum model, we calculate the two-particle binding energy $\varepsilon_b$ in the field theory, and  using the parameters $m$ and $g$ given by Eq.~\eqref{couplingconstant} and~\eqref{eq:CoefficientSchriefferModel}, compare it with the exact solution of the lattice model. The expression for $\varepsilon_b$ is 
\begin{eqnarray} \label{eq:ContBinding} 
\frac{\varepsilon_b}{\varepsilon_{\rm uv}} &=& \left[ e^{1/g_0} -1 \right]^{-1},  \nonumber \\
\quad g_0 &=& \frac{9}{\pi} \frac{2\lambda-V_f}{W} = \frac{6}{\pi} \frac{V^2}{\Delta (\Delta +2V)} \, ,
\end{eqnarray}
with $\varepsilon_{\rm uv} = \pi W / 9$ an energy cutoff that we fix with the exact binding energy at $V \to \infty$~\cite{SuppMat}. 
The exponent $g_0$ in Eq.~\eqref{eq:ContBinding}, defined by the ratio of effective pairing interaction $g$ and the bandwidth, only depends on the ratio $V/\Delta$ in the narrow band regime $t\ll \Delta$.   
Remarkably, a perfect agreement between continuum theory and lattice model is found at all values of $V/\Delta$ (see Fig.~\ref{fig:PairBindingBandwidth}). This proves the accuracy of our mapping from lattice model to continuum theory.  

Let us summarize our achievements so far. We have transformed the strongly repulsive model (Eq.~\ref{eq:OriginalModel}), into an effective Hamiltonian for doped particles featuring attractive interaction. When the single particle bandwidth is small, this transformation is exact and the two-particle pairing problem is solved exactly. The nature of the ground state at low density depends on the strength of this attraction, measured by the dimensionless coupling constant $g_0$. If $g_0$ is small, the binding energy is small compared to the Fermi energy and the doped charge form a weakly attractive Fermi gas. On the other hand, a large $g_0$ will produce tightly bound pairs. One can tune between these two regimes by increasing the ratio $V/\Delta$, as shown Fig.~\ref{fig:PairBindingBandwidth}a.

\begin{figure}
\centering
\includegraphics[width=\linewidth]{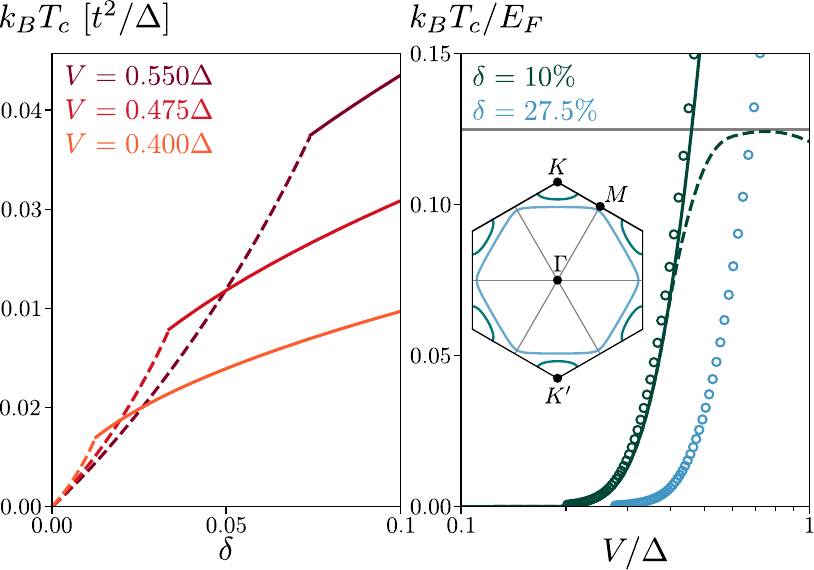}
\caption{\textbf{Critical temperature:} Critical temperature of the continuum model in the exact BCS (solid, Eq.~\eqref{eq:BCSCriticalTemp}) and BEC (dashed) limits, as a function of doping and $V/\Delta$. A gray horizontal line highlights the bound $k_B T_c < E_F/8$. The self-consistent lattice mean-field solutions (dots) agree with the continuum theory for small doping, but differ when the $\pm K$ pockets merge (inset).  
}
\label{fig:CriticalTemperature}
\end{figure}

The attractive Fermi gas in two dimensions is known to be superconducting at low temperature and exhibits a BCS-BEC crossover as the ratio between pair binding and Fermi energies changes from small to large values~\cite{randeria1989bound,bertaina2011bcs,parish2015bcs}. 
In the region of weakly bound pairs $\varepsilon_b \ll E_F$, the critical temperature is given by $k_B T_c = e^{\gamma-1} \sqrt{ 2 E_F \varepsilon_b} / \pi$ with $\gamma \simeq 0.577$ Euler's constant~\cite{miyake1983fermi,petrov2003superfluid}. In terms of the dimensionless coupling constant $g_0$ (Eq.~\eqref{eq:ContBinding}), we obtain the explicit formula for $T_c$: 
\begin{equation} \label{eq:BCSCriticalTemp}
k_B T_c = e^{\gamma-1} \sqrt{ \frac{2E_F W}{ 9\pi  }} e^{-1/(2g_0)}   \, ,
\end{equation}
where the Gorkov Melik-Barkhudarov corrections have been included to correctly describe the strong coupling nature of the superconducting state~\cite{gor1961contribution,chubukov2016superconductivity}. 
As an example, this formula safely applies at $\delta = 0.1$ if $V < 0.7 \Delta$, where we both have $\exp(1/g_0) > 10$ and $\varepsilon_b \lesssim E_F/2$~\cite{bertaina2011bcs}, as confirmed thereafter by exact diagonalization.

On the other side of the crossover $\varepsilon_b \gg E_F$, the physics depends  on the interaction between the bosonic pairs. When these bosons repel, the system exhibits a BKT transition toward a BEC at low temperature~\cite{fisher1988dilute}, while it collapses if bosons attract~\cite{ruprecht1995time,roberts2001controlled}. Between the extreme BCS and BEC limits, the critical temperature satisfies the very general bound $k_B T_c \leq E_F / 8$ ~\cite{hazra2019bounds}, which limits the largest achievable $T_c$.

The BCS-BEC crossover of the two-dimension Fermi gas can be achieved by tuning either carrier density or the interaction strength $g_0$, which is controlled by $V/\Delta$ in our model.  We plot in Fig.~\ref{fig:CriticalTemperature} the critical temperature $T_c$ as a function of doping concentration $\delta$ and $V/\Delta$. At very low doping where $E_F < \varepsilon_b$, the system lies in the BEC regime and $T_c$ increases rapidly with $\delta$. At some critical concentration, the system undergoes the BEC-BCS crossover and finally follows Eq.~\eqref{eq:BCSCriticalTemp}.

It is worth emphasizing that our exact BCS formula Eq.~\eqref{eq:BCSCriticalTemp} applies provided that the dimensionless coupling constant $g_0$ is small, even when the bare repulsion $V$ far exceeds the bandwidth $W$. 
This is because doped fermions at the conduction band bottom $\pm K$ reside entirely on $B$ sublattice, and therefore avoid the direct nearest-neighbor repulsion $V$. In the weak-coupling regime $V\ll \Delta$, the attraction $g_0 \propto (V/\Delta)^2$ between low-energy carriers is induced by virtual interband particle-hole pairs or excitons, and leads to exponentially small $T_c$.    
Most importantly however, our expression of $g_0$ is nonperturbative in $V$ and remains exact at $V \sim \Delta \gg W$, where strong-coupling superconductivity and maximum $T_c$ are attained.  
For instance, at doping  $\delta=0.1$, $T_c$ reaches $0.1E_F \simeq 0.032 t^2/\Delta$ around $V=0.43 \Delta$, which is about $0.5\%$ of the quasiparticle bandwidth $W=9t_f \simeq 6.3 t^2/\Delta$.

\begin{figure}
\centering
\includegraphics[width=\columnwidth]{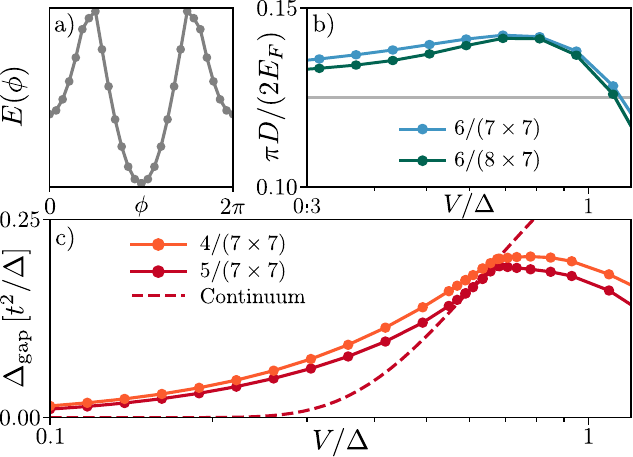}
\caption{ \textbf{Numerical evidence of superconductivity:}
a) The ground state energy exhibits the characteristic flux dependence of a superconductor, as shown here for $V = \Delta$ and 6 particles on a $8\times 6$ lattice.
b )The charge stiffness divided by the Fermi energy is constant in the BCS limit, reaching 1/8 for small doping (gray). 
c) Superconducting gap as a function of $V/\Delta$ for different doping concentrations. It follows the continuum prediction (dashed) up to $V/\Delta \simeq 0.7$.
}
\label{fig:ChargeStiffnessNumerics}
\end{figure}

\section*{Materials and Methods}

We support the emergence of superconductivity with evidence from ED on finite size lattices.
While the original model Eq.~\eqref{eq:OriginalModel} exhibits positive $\varepsilon_b$~\cite{SuppMat}, we focus on the effective model Eq.~\eqref{eq:BelectronEffModel} which allows to reach larger system sizes. 
First, the superfluid behavior of the system is probed by the charge stiffness~\cite{Kohn_MetalInsulating}
\begin{equation} \label{eq:DefChargeStiffness}
D = \frac{1}{16\pi^2} \left. \frac{L_1}{L_2} \frac{\partial^2 E(N, \phi)}{\partial \phi^2} \right|_{\phi=0}
\end{equation}
at doping $\delta = N /(L_1 \times L_2)$, with $N$ the number of doped fermions and $(L_1,L_2)$ the number of sites along the two basis vectors of the triangular lattice. 
$D$ measures the sensitivity of the ground state energy $E(N, \phi)$ to twisted boundary conditions $\psi_{r+L_1} = e^{2 i \pi \phi} \psi_{r}$. 
A positive value of $D>0$ in the thermodynamic limit implies dissipationless charge transport and gives a direct signature of the Meissner effect~\cite{scalapino1993insulator}. 
In the range of parameters considered, our system clearly exhibits (with small finite-size effect) the $h/2e$ flux periodicity of superconductors~\cite{FluxPeriodicity_SC} and shows positive $D$ (see Fig.~\ref{fig:ChargeStiffnessNumerics}a-b), which proves superconductivity in the ground state.
The charge stiffness of a BCS superconductor with a parabolic dispersion relation is known exactly: $D = E_F/4\pi$ ~\cite{hazra2019bounds}. 
Our results for $V < \Delta$ are correctly captured by this prediction, especially at low doping concentrations where fermions live close to the band minima.

To demonstrate the strong-coupling nature of the superconducting state, we  consider the superconducting gap
\begin{equation}
\Delta_{\rm gap} = \frac{(-1)^N}{2} \left[ E(N+1) + E(N-1) - 2 E(N) \right] \, .
\end{equation}
Our continuum theory Eq.~\eqref{eq:ContinuumModel} predicts $\Delta_{\rm gap} =\sqrt{2E_F \varepsilon_b}$ up to $V \sim \Delta$. This leads to a ratio of the gap and critical temperature $\Delta_{\rm gap} / k_B T_c = \pi e^{1-\gamma} \simeq 4.796$~\cite{miyake1983fermi,petrov2003superfluid,gor1961contribution}, which is much larger than the universal value $1.764$ in BCS theory for weak-coupling superconductors~\cite{schrieffer2018theory}. 
This is because  the phonon-induced retarded attraction in conventional metals is limited to electrons within a Debye energy from the Fermi surface, whereas in our theory the induced pairing interaction is instantaneous on the time scale of inverse bandwidth ($\hbar/W$), so that all carriers in the narrow band are subject to the pairing interaction. For sufficiently large $V$ or at very small doping concentrations, the system lies in the BEC regime and the gap to $T_c$ ratio can take arbitrarily large values.

Our numerical results for $\delta \simeq 0.1$, shown in Fig.~\ref{fig:ChargeStiffnessNumerics}c, confirm the superconducting behaviors identified above with a robust gap $\Delta_{\rm gap}$ increasing with $V$ up to $V/\Delta = 0.7$. Near this point, our numerical results agree with the mean field prediction, and the gap reaches as large as $\Delta_{\rm gap} = 0.84 E_F$. This allows the system to reach critical temperature of 0.1$E_F$, as described above. 
For $V \lesssim 0.5 \Delta$, the numerically extracted gaps $\Delta_{\rm gap}$ lie above the continuum theory prediction due to finite size effects. Indeed in that regime, the finite lattice considered cannot fully accommodate the large-sized bound state that arise in the thermodynamic limit (see Fig.~\ref{fig:PairBindingBandwidth}). This effective confinement increases the energy of the bound pairs, resulting in an overestimate of  the superconducting gap. Despite this discrepancy, the simultaneous presence of a positive $\Delta_{\rm gap}$ and a non-zero charge stiffness stands as a strong probe of superconductivity in our model for $V < 0.7 \Delta$.

In addition to the BCS-BEC superconductivity at low density, our model shows very rich physics at higher doping concentrations, where lattice effects become important. 
By performing mean-field calculation on the model Eq.~\eqref{eq:BelectronEffModel}, we find $f$-wave pairing for doping $\delta < 1/3$ (see supplementary materials~\cite{SuppMat}). 
The corresponding critical temperatures, shown in Fig.~\ref{fig:CriticalTemperature}, are calculated with the linearized gap equation
\begin{equation}
\frac{1}{\alpha} = \frac{1}{N_s} \sum_{\vec{q}} \frac{ \left[ \sum_j \sin(\vec{q} \cdot \vec{a}_j)  \right]^2  }{|\xi_{\vec{k}}|}\tanh \left(\frac{|\xi_{\vec{k}}|}{2k_B T_c}\right)
\end{equation}
with $\xi_{\vec{k}}=\varepsilon (\vec{k})-\mu$ and $\alpha = 2\lambda - V_f - (2 \beta +\delta) U_3$, where $\beta  =  (3 N_s)^{-1} \sum_{\vec{q}} f_{\rm FD}  (\xi_{\vec{k}}) \sum_j \cos(\vec{q}\cdot \vec{a}_j)$ originates from three-body interactions. Here, $N_s$ denotes the total number of sites, and 
the chemical potential is fixed by $\delta = N_s^{-1} \sum_{\vec{q}} f_{\rm FD} (\xi_{\vec{k}})$,
with $f_{\rm FD}$ the Fermi-Dirac distribution.

At low doping, the $f$-wave superconducting state has a full pairing gap and its $T_c$  obtained from lattice model calculation agrees well with our previous result based on continuum theory. 
Interestingly, at higher doping concentration $\delta> \delta_c \approx 1/4$, the gap vanishes at 6 nodes on the Fermi surface along $\Gamma M$ direction, where the $f$-wave gap function vanishes. This change in gap structure is due to the change of Fermi surface topology across the van Hove singularity, where the two pockets around $\pm K$ merge into a single Fermi surface enclosing the $\Gamma$ point (see Fig.~\ref{fig:CriticalTemperature}).

The above conclusions are confirmed by our ED study, which shows clear evidence of nodal superconductivity at $\frac{1}{4}<\delta\leq \frac{1}{3}$. 
We also find that the superconducting state is remarkably robust against longer range bare repulsion. These numerical results can respectively be found in the supplementary materials~\cite{SuppMat}.

Finally, our ED study reveals non-superconducting states in the ultrastrong coupling regime $V \gtrsim \Delta$~\cite{SuppMat}.  The detailed description of these competing phases is left for further study, the focus of this work being the fully-controlled theory of superconductivity emerging from repulsive interactions at $V < \Delta$. 

\section*{Discussion}

Our work opens a new route to unconventional superconductivity in atomic Fermi gas and electron systems. Encouragingly, optical lattice with honeycomb geometry and tunable band gap have already been realized~\cite{uehlinger2013artificial,flaschner2016experimental}.  Many recent advances in dipolar or Rydberg atom systems with longer-range interactions~\cite{baier2018realization,lu2012quantum} have enabled the implementation of one-dimensional $t-V$ Hamiltonian~\cite{guardado2020quench}, and hold great promise for the realization of our model in near future.

Our mechanism for strong-coupling superconductivity mediated by charge-transfer complex, or interband excitations, may also shed insight on graphene-based moir\'e superlattices, where the small bandwidth and high $k_B T_c / E_F$ ratio (up to $\sim 0.1$) place important constrains on viable theories. It will be interesting to develop accurate low-energy models for these systems and analyze superconductivity in the narrow band limit as exemplified in our work. 
In this regard, we note that correlated hopping and direct repulsion also appear in effective Hamiltonian for narrow bands in twisted bilayer graphene~\cite{guinea2018electrostatic,KangVafek,PhysRevX.8.031087}. Our theory suggests that renormalization by virtual interband excitations is necessary to obtain strong-coupling superconductivity.   
Moreover, our simple model may be relevant to twisted double bilayer graphene~\cite{liu2020tunable} and trilayer graphene-boron nitride heterostructures~\cite{chen2019signatures}, where signs of spin-polarized superconductivity has been reported~\cite{lee2019theory,cornfeld2020spin}. We leave to future work the extension of our theory to spinful systems and its application to various strongly-correlated materials.

{\it Acknowledgments. } We thank K. Slagle and A. Chubukov for valuable comments on the manuscripts. This work was supported by DOE Office of Basic Energy Sciences, Division of Materials Sciences and Engineering under Award DE-SC0018945. LF was supported in part by a Simons Investigator Award from the Simons Foundation.

{\it Author contributions.} Both authors contributed essentially to the formulation and theoretical analysis of the problem and to writing the manuscript. VC performed numerical calculations. 

{\it Statement on competing interests.} All authors declare that they have no competing interests. 

{\it Data and materials availability.} All data needed to evaluate the conclusions in the paper are present in the paper and/or the Supplementary Materials. 

\bibliography{Biblio_TrimerSC}

\onecolumngrid
\newpage
\begin{center}
\textbf{\large Supplemental Materials: New mechanism and exact theory of superconductivity from strong repulsive interaction}\\[10pt]
Valentin  Cr\'epel,  Liang  Fu \\
\textit{Massachusetts Institute of Technology, 77 Massachusetts Avenue, Cambridge, MA, USA}
\end{center}
\twocolumngrid

\renewcommand{\thefigure}{S\arabic{figure}} 
\renewcommand{\thesection}{S\arabic{section}} 
\renewcommand{\theequation}{S\arabic{equation}}
\setcounter{figure}{0}
\setcounter{section}{0}
\setcounter{equation}{0}

\section{Schrieffer-Wolff Transformation} \label{app:SchriefferWolff}

In this appendix, we derive effective model Eq.~\ref{eq:BelectronEffModel} with a Schrieffer-Wolff transformation. This effective Hamiltonian is exact up to corrections of order $(t/\Delta)^2$, irrespective of the ratio $t/V$. 

\subsection{Canonical Transformation}

Single particle tunneling on the honeycomb lattice couples the $f$-band to trimers, polarons and dipoles. 
The Schrieffer-Wolff transformation uses a canonical transformation $\mathcal{H}' = e^{iS} \mathcal{H} e^{-iS}$, with $S$ Hermitian, to treat these couplings as an effective Hamiltonian that leaves the $f$-band invariant up to second order correction in $t/\Delta$. 
This is achieved if $S$ satisfies 
\begin{equation} \label{appeq:Schrieffer_Condition}
[\mathcal{H}_0, iS] =  \mathcal{H}_t \, ,
\end{equation}
as can be seen with the Baker-Campbell-Haussdorf formula. 
Under the assumption Eq.~\ref{appeq:Schrieffer_Condition}, we get
\begin{equation} \label{appeq:Schrieffer_SecondOrderGeneric}
\mathcal{H}' = \mathcal{H}_0 + \frac{1}{2} \left[ iS, \mathcal{H}_t \right] + \mathcal{O} ( \mathcal{H}_t S^2) \, . 
\end{equation}

To find $S$, we follow Ref.~\cite{MacDonald_tOverVExpansionHubbard} and split the tunneling Hamiltonian into a collection of operators $T_{p,M}$
\begin{equation}
\mathcal{H}_t = \sum_{p=\pm 1} \sum_{M = -2}^2 T_{p,M} \, ,
\end{equation}
where $T_{p,M}$ gathers all tunneling operations that change the number of occupied $A$ sites by $(-p)$ and the number of occupied nearest neighbor pairs by $M$, \textit{i.e.} it changes the energy of $\mathcal{H}_0$'s eigenstates by $p\Delta + MV$. The bounds for $p$ and $M$ are determined by the lattice geometry. 
With these definition, we can check that
\begin{equation} \label{appeq:Schrieffer_ExpressionOfS}
S = -i \sum_{p, M} \frac{T_{p,M}}{M V + p \Delta} \, , 
\end{equation}
satisfies Eq.~\ref{appeq:Schrieffer_Condition}.
Consider an eigenstate $\ket{n}$ of $\mathcal{H}_0$ with energy $E_n$. 
By definition $T_{p,M} \ket{n}$ is also an eigenstate of $\mathcal{H}_0$ with energy $E_n + (p\Delta + MV)$. As a consequence, we have
\begin{align}
[\mathcal{H}_0, iS] \ket{n} & = \sum_{p,M} \frac{(E_n + p\Delta + VM) T_{p,m} - T_{p,M} E_n }{p\Delta + VM}\ket{n} \notag \\ & = \sum_{p,M}T_{p,M}\ket{n} = \mathcal{H}_t \ket{n} \, .
\end{align}
Finally, the relation $T_{p,M}^\dagger = T_{-p, -M}$ ensures the hermiticity of $S$. 
Plugging Eq.~\ref{appeq:Schrieffer_ExpressionOfS} into Eq.~\ref{appeq:Schrieffer_SecondOrderGeneric}, we find the generic expression
\begin{equation} \label{appeq:Schrieffer_SecondOrderGeneric2}
\mathcal{H}' = \mathcal{H}_0 + \frac{1}{2} \sum_{p,p', M, M'} \frac{[T_{p',M'}, T_{p,M}]}{M' V +  p' \Delta} + \mathcal{O} \left( \frac{t^3}{\Delta^2} \right) \, .
\end{equation}
To evaluate the correcting terms of this last equation, we have noticed that all terms in $S$ are smaller than $t/\Delta$ in magnitude.

\subsection{Low energy projection}

To describe the low-energy properties of our model, we project the obtained Hamiltonian onto the $f$-band which is well separated from other excitations by the charge transfer gap $\Delta \gg t$. 
This projection restricts the sum in Eq.~\ref{appeq:Schrieffer_SecondOrderGeneric2} to cases where $p'=-p$ and $M'=-M$. 
Furthermore, the first operator acting on the states of the $f$-band should move an electron from an $A$ to a $B$ site, \textit{i.e.} the rightmost $T_{p,M}$ must have $p,M \geq 0$. 
This gives 
\begin{equation} \label{appeq:Schrieffer_SecondOrderGeneric3}
\mathcal{H}' \simeq \mathcal{H}_0 - \sum_{M = 0, 1, 2} \frac{T_{-1, -M} T_{1,M}}{MV+\Delta} \, .
\end{equation}
The three terms of the sum $M=0,1,2$ gather all second order processes that respectively involve the virtual occupation of a trimer, a polaron and a dipole excitation (shown in Fig.~\ref{fig:HoneycombLattice}).
As expected from second order perturbation theory, these processes occur with rates
\begin{equation} \label{appeq:Schrieffer_Rates}
t_T = \frac{t^2}{\Delta}  , \quad t_P = \frac{t^2}{V+\Delta}  , \quad \text{and } \quad t_D = \frac{t^2}{2V+\Delta}  ,
\end{equation}
that are inversely proportional to their energy difference with the $f$-band.

\subsection{Simplification}

To further simplify $\mathcal{H}'$, we isolate the contributions of Eq.~\ref{appeq:Schrieffer_SecondOrderGeneric3} depending on the number of occupied neighbor of each $A$-site.
In the following, we focus on a $A$-site at position $r_0$ and denote its three neighbors as $r_1$, $r_2$ and $r_3$. 

\paragraph{No neighboring fermions:} If the $A$-site is surrounded by three empty sites, it can couples to three different dipole excitations in which the fermion at $r_0$ is moved on a $B$ neighboring site. This provides an energy shift
\begin{equation} \label{appeq:DipoleContrib}
H_{0f} = - 3 t_D (1-n_{r_1}) (1-n_{r_1}) (1-n_{r_1}) \, .
\end{equation}

\paragraph{One neighboring fermion:} If the $A$-site has a neighboring fermion at position $r_1$, it can couples to two polarons by hopping to either $r_2$ or $r_3$. 
If the second tunneling process in Eq.~\ref{appeq:Schrieffer_SecondOrderGeneric3} moves it back to $r_0$, the process leads to an energy shift. 
On the contrary, if the fermion at $r_1$ replace the original one at $r_0$, the process can be viewed as an effective tunneling of the $r_1$-fermion on a neighboring $B$-site. Together, they give 
\begin{equation}\label{appeq:PolaronContrib} \begin{split}
H_{1f} =  - & 2 t_P  n_{r_1} (1-n_{r_2}) (1-n_{r_3})  \\ & + t_P [ f_{r_2}^\dagger (1-n_{r_3}) + f_{r_3}^\dagger (1-n_{r_2}) ] f_{r_1} \, .
\end{split} \end{equation}
Note that the tunneling terms come with a positive sign due to the anticommutation relation between fermions. Circular permutation of the indices $(r_1,r_2,r_3)$ gives all other possible terms involving a polaron.

\paragraph{Two neighboring fermions:} Finally, if $r_1$ and $r_2$ both host a fermion, the system can only couple to a trimer excitation by moving the $A$ electron to $r_3$. The second tunneling process can either put this fermion back to $r_0$, leading to en energy reduction, or put the ones at $r_1$ or $r_2$ at $r_0$, giving a density-assisted tunneling term:
\begin{equation} \label{appeq:TrimerContrib}
H_{2f} = - t_T  n_{r_1} n_{r_2} (1-n_{r_3})  + t_T ( f_{r_2}^\dagger n_{r_3} + f_{r_3}^\dagger n_{r_2} )  f_{r_1} \, .
\end{equation}
Again, circular permutation of the indices $(r_1,r_2,r_3)$ gives all other possible terms involving a trimer. 

\begin{figure}
\centering
\includegraphics[width=\columnwidth]{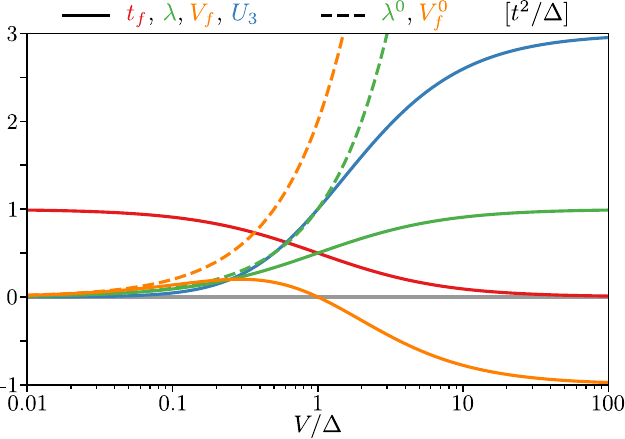}
\caption{\textbf{Coefficients of the effective model:}
Coefficients of the effective model Eq.~\ref{eq:BelectronEffModel} (full lines), and projection of the bare repulsion into the $f$-band (dashed) as a function of $V/\Delta$.
}
\label{figapp:SchriefferCoeffs}
\end{figure}

Gathering the contributions of \textit{a-b-c}, we obtain the effective Hamiltonian Eq.~\ref{eq:BelectronEffModel} where the coefficients originates from the processes outlined in Fig.~\ref{fig:HoneycombLattice}. 
The single particle fermion tunneling only appears in Eq.~\ref{appeq:PolaronContrib} and therefore reads $t_f = t_P$. 
The correlated tunneling appears in Eqs.~\ref{appeq:TrimerContrib} and~\ref{appeq:PolaronContrib} with opposite signs: $\lambda = t_T - t_P$.
Two-body and three-body interaction coefficients come from the three cases \textit{a-b-c}. Carefully counting all terms, we find $V_f = -t_T + 4 t_P - 3t_D$ and $U_3 = 3t_T - 6t_P + 3 t_D$. 
These expression are given in Eq.~\ref{eq:CoefficientSchriefferModel}, and plotted against $V/\Delta$ in Fig.~\ref{figapp:SchriefferCoeffs}. For comparison (see text), we also show the results from the projection of the bare repulsion $V$ into the $f$-band, valid when $V \ll \Delta$.

\section{Solution of the two body problem} \label{app:TwoBodyProblem}

In this appendix, we solve the effective lattice model Eq.~\ref{eq:BelectronEffModel} for two fermions. To understand the competing roles of two-body interaction and correlated hopping, we first look at the quartic part of the Hamiltonian (Sec.~\ref{sapp:ResonatingTriangle}). Then, we decouple the center of mass and relative motions of the two fermions (Sec.~\ref{sapp:COMvsRelative}) and explain in more detail how the results presented in the main text were obtained (Sec.~\ref{sapp:FiniteSizeScaling}). 

\subsection{'Resonating' triangle} \label{sapp:ResonatingTriangle}

\begin{figure}
\centering
\includegraphics[width=\columnwidth]{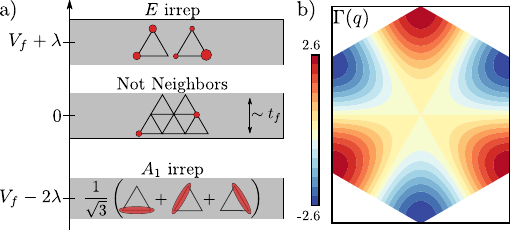}
\caption{\textbf{Bound states as resonating triangles:}
a) The lowest energy subspace of $\mathcal{H}_{\rm int}'$ is composed of bound states ($E_{A_1} = V_f - 2\lambda <0$), which can be seen as dimers 'resonating' within an upper triangle of the $B$-lattice. b) Weight of the two-body ground state $\Gamma(q)$ in the limit $V\gg \Delta$. It is maximal in amplitude and opposite in sign at the $K$ and $K'$ points, underlying its valley-singlet nature.
}
\label{figapp:ResonatingTriangle}
\end{figure}

We start with the quadratic terms of our model
\begin{equation}
\mathcal{H}_{\rm int}' = V_f \sum_{\langle i, j \rangle} n_i n_j + \lambda \sum_{ (ijk) \in \triangle } (f_i^\dagger n_j f_k + P_{ijk}) \, .
\end{equation}
All states where the two electrons are not neighbors are zero energy eigenstates of this operator. The others have two doped charge on the same upper triangle, and can be divided into irreducible representations of $C_{3v}$. More precisely, there is a state in the identity representation $A_1$ and an $E$-doublet
\begin{align}
&\ket{A_1, \vec{r}} = \frac{1}{\sqrt{3}}  \left[ f_{\vec{r}}^\dagger f_{\vec{r}+\vec{a}_1}^\dagger + f_{\vec{r}+\vec{a}_1}^\dagger f_{\vec{r}-\vec{a}_3}^\dagger + f_{\vec{r}-\vec{a}_3}^\dagger f_r^\dagger\right] \ket{n=1} \notag \\
&\ket{E, \vec{r}} = \frac{1}{\sqrt{6}}  \left[ f_{\vec{r}}^\dagger f_{\vec{r}+\vec{a}_1}^\dagger + f_{\vec{r}+\vec{a}_1}^\dagger f_{\vec{r}-\vec{a}_3}^\dagger -2 f_{\vec{r}-\vec{a}_3}^\dagger f_r^\dagger\right] \ket{n=1} \notag \\
&\ket{E', \vec{r}} = \frac{1}{\sqrt{2}}  \left[ f_{\vec{r}}^\dagger f_{\vec{r}+\vec{a}_1}^\dagger - f_{\vec{r}+\vec{a}_1}^\dagger f_{\vec{r}-\vec{a}_3}^\dagger \right] \ket{n=1} \, ,
\end{align}
with $\ket{n=1}$ the insulating state at unit filling, which also corresponds to the vacuum for the $f$ operators. These three eigenstates of $\mathcal{H}_{\rm int}'$ have energy
\begin{equation}
E_{A_1} = V_f - 2 \lambda \, , \quad E_E = V_f + \lambda \, .
\end{equation}
Because $V_f - 2 \lambda < 0$, the lowest energy manifold is made of all states $\{ \ket{A_1, \vec{r}} \}_{r \in B}$ and has \emph{negative} energy, corresponding to bound states. As mentioned in the main text, this binding energy solely comes from correlated hopping when $V \leq \Delta$, for which $V_f > 0$. These dimers have the $f$-wave symmetry described in the main text (see Fig.~\ref{fig:PairBindingBandwidth}), and can be seen as dimers 'resonating' within an upper triangle as shown in Fig.~\ref{figapp:ResonatingTriangle}. 

This solution unveils the crucial role of correlated tunneling for pairing and gives an intuitive real-space picture for the bound pairs. It also allows to better understand the low energy continuum model Eq.~\ref{eq:ContinuumModel}. First, the coupling constant $g$ is directly proportional to the energy $E_{A_1}<0$. Moreover, it also gives a physical intuition for the $s$-wave scattering between the two valleys. To see this, let's reintroduce the tunneling $t_f$ in the lowest energy subspace $\{ \ket{A_1, \vec{r}} \}_{r \in B}$ to lift the degeneracy of the low energy manifold (this perturbative treatment of $t_f$ is justified in the limit $V \gg \Delta$). The dispersion relation of the bound pairs $\varepsilon_{A_1} (k) = E_{A_1} - (2 t_f/3) \sum_{j=1}^3 \cos \left( \vec{k} \cdot \vec{a}_j \right)$ has its minimum at the $\Gamma$ point. The lowest energy state is
\begin{equation} \label{eq:LowestEnergyStateGammaPoint}
\ket{\Gamma^{2e}} = \frac{i}{\sqrt{3N_s}} \sum_{\vec{q} \in BZ} \Gamma(\vec{q})  f_{-\vec{q}}^\dagger f_{\vec{q}}^\dagger \ket{n=1} \, ,
\end{equation}
with $\Gamma(\vec{q}) =  \sum_{j=1}^3 \sin(\vec{q} \cdot \vec{a}_j )$. The amplitude of the paired state $\Gamma(q)$ is maximal in magnitude near the $K$ and $K'$ points, as shown in Fig.~\ref{figapp:ResonatingTriangle}. This confirms our intuition that the dominant scattering channel couples electrons with different isospin $\tau = \pm$ in $s$-wave to form valley singlets. While this argument gives an intuitive understanding of the dominant pairing interaction in our system, it can only be formally applied in the limit where $V \gg \Delta$ where the tunneling $t_f$ can be treated as a perturbation to $\mathcal{H}_{\rm int}'$.

\subsection{Center of mass and relative motion} \label{sapp:COMvsRelative}

To solve the two-particle problem exactly, we now decouple the center of mass and relative motion of the two fermions. Taking all terms of Eq.~\ref{eq:BelectronEffModel} simultaneously into account, we obtain the results presented in Fig.~\ref{fig:PairBindingBandwidth}, which also agree with the previous perturbative treatment of the previous section in the limit $V \gg \Delta$.

The Hilbert space with two fermions is spanned by the states 
\begin{equation}
\ket{r_1, r_2} = f_{r_1}^\dagger f_{r_2}^\dagger \ket{n=1} \, .
\end{equation}
Taking advantage of the translation invariance of the problem, we can introduce the center of mass momentum $K$ and reorganize the Hilbert space with the Bloch-waves~\cite{vidal2000interaction}
\begin{equation}
\ket{\varphi(K, r)} = \frac{1}{\sqrt{N_s}} \sum_R e^{i(K\cdot R)} \ket{R, R+r} \, .
\end{equation}
Because $\ket{r_2, r_1} = -\ket{r_1, r_2}$, the state with opposite relative positions $r$ describe the same physical state $\ket{\varphi(K,-r)} = - e^{i (K \cdot r)} \ket{\varphi(K,r)}$. To avoid double counting the states, we restrict our attentions to states with $(r \cdot \delta_1) \geq 0$. The action of the Hamiltonian in this new basis can be directly computed
\begin{align} 
& \mathcal{H}' \ket{\varphi(K, r)} = \\
& \sum_{\substack{j=1,2,3 \\ \epsilon = \pm}} t_f [ 1 + e^{i \epsilon (K\cdot a_j)} ] \ket{\varphi(K, r + \epsilon a_j)}  + V_f  \delta_{r,\epsilon a_j}  \ket{\varphi(K, r)}  \notag \\
& + \lambda \sum_{j} \delta_{r,a_j} \ket{\varphi(K, -a_{j-1} )} + \delta_{r,-a_{j-1}} \ket{\varphi(K, a_j )} \notag \\
& + \lambda \sum_{j} e^{-i(K\cdot a_{j+1})} \delta_{r,-a_j} \ket{\varphi(K, a_{j-1} )} \notag \\
& + \lambda \sum_{j} e^{i(K\cdot a_{j-1})} \delta_{r,a_j} \ket{\varphi(K,-a_{j+1} )} \, . \notag 
 \end{align} 
As expected from translation invariance, the state $\ket{\varphi(K, r)}$ only couple to states with the same center of mass momentum $K$. The tunneling on the lattice acts as an effective hopping term for the relative position $r$. The local interaction and correlated tunneling terms only changes this tight-binding Hamiltonian close to the origin for $r=a_j$ or $r=-a_j$ with $j=1,2,3$. This tight binding Hamiltonian can be solved numerically for each center of mass momentum $K$. Its ground state always is at the $\Gamma$ point, and reads $\ket{\psi_{\rm bd}} = \sum_r \psi_{\rm bd}(r) \ket{\varphi(0, r)}$. Its energy yields the binding energy and bound state size
\begin{equation}
\xi_{\rm bd} = \sqrt{ \sum_r r^2 |\psi_{\rm bd}(r)|^2 } \, , 
\end{equation} 
in Fig.~\ref{fig:PairBindingBandwidth}. Of course, accurate results can only be obtained when the system size exceeds the spread of the two-fermion bound state. In particular for $V/\Delta<0.25$, the bound state size sharply increases and a more careful extraction of the energy is required.

\subsection{Finite size scaling} \label{sapp:FiniteSizeScaling}

\begin{figure}[t!]
\centering
\includegraphics[width=\columnwidth]{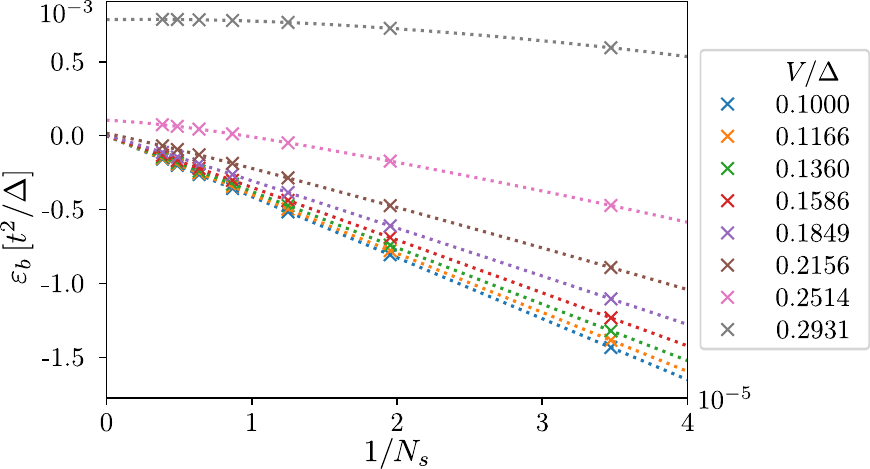}
\caption{ \textbf{Binding energy as a function of the inverse number of site:} The linear behavior is extrapolated to $1/N_s \to 0$ for $V/\Delta<0.25$ where extremely large system sizes are needed to accommodate the very weakly bound states (see Fig.~\ref{fig:PairBindingBandwidth}).  
}
\label{figapp:FiniteSizeScaling}
\end{figure}

We therefore focus on the regime $V/\Delta<0.25$, and extract the two-body problem as a function of the inverse number of site in the system $1/N_s$ as shown in Fig.~\ref{figapp:FiniteSizeScaling}. Linear interapolation of these curves at $1/N_s=0$ gives the values of the binding energy in the thermodynamic limit. The interpolated values are reported in Fig.~\ref{fig:PairBindingBandwidth}. It is worth pointing out that we infer positive pair binding energy in the entire range $V/\Delta>0$, in agreement with the continuum model.

\section{Continuum model} \label{app:ContinuumModel}

\subsection{Derivation}

In this appendix, we derive the continuum model Eq.~\ref{eq:ContinuumModel} in the limit of small doping concentration. Focusing on the physics near the $K$ and $K'$ pockets, and discarding all high energy fermionic degrees of freedom amounts to write the Fourier transform of the $f$-operators as
\begin{equation}
f_r = \frac{1}{\sqrt{N_s}} \sum_{\tau = \pm} \sum_{k, \, ka \ll 1} e^{-i [(\tau K + k)\cdot r]} \psi_{\tau, k} \, ,
\end{equation}
with $N_s$ the number of honeycomb unit cells. Total momentum conservation restricts the interaction terms to valley-preserving ones. They can be further decomposed as intra-valley interactions 
\begin{equation}
H_p' = \frac{1}{2 N_s} \sum_{k,q,q'} \left[ \sum_{\tau} V_{k,q+\tau K} \psi_{\tau, q-k}^\dagger \psi_{\tau, q'+k}^\dagger \psi_{\tau, q'} \psi_{\tau, q} \right]
\end{equation}
and inter-valley interactions 
\begin{widetext} \begin{equation}
H_s' = \frac{1}{2 N_s} \sum_{k,q,q'}  \left( V_{k,q+K} + V_{-k,q'-K} - V_{q'-q+k-2K,q'-K} - V_{q-q'-k+2K,q+K}  \right) \psi_{+, q-k}^\dagger \psi_{-, q'+k}^\dagger \psi_{-, q'} \psi_{+, q} \, ,
\end{equation} \end{widetext}
which both depend on the momentum space representation of the quartic terms of Eq.~\ref{eq:BelectronEffModel}:
\begin{equation} \label{appeq:InteractionMomentumSpace}
V_{k,q} = \frac{V_f}{2 t_f} \varepsilon (k) + \lambda \sum_{j=1}^3 e^{i(q\cdot a_j + k \cdot a_{j+1})} + e^{-i(q\cdot a_j + k \cdot a_{j-1})} .
\end{equation}
Due to the small momenta considered in our continuum theory, we can Taylor expand the previous expression with the help of the relations
\begin{equation}
V_{k, q+\tau K} = 3 (V_f - \lambda)  , \, V_{k+2\tau K, q+\tau K}  = -\frac{3}{2} V_f + 6\lambda  ,
\end{equation}
which hold true up to $\mathcal{O}(k,q)$ corrections. 
This leading order approximation and the fermion anti-commutation relations show that intra-valley interactions $H_p' \simeq 0$ are negligible compared to inter-valley ones
\begin{equation}
H_s' = \frac{V_0}{2N_s} \sum_{k,q,q'}  \psi_{+, q-k}^\dagger \psi_{-, q'+k}^\dagger \psi_{-, q'} \psi_{+, q} \, , 
\end{equation}
with $V_0 =  9 (V_f - 2 \lambda) = -9U_3$. 
Fourier transforming back to real-space and accounting for the Brillouin zone area, we obtain the effective interaction announced in Eq.~\ref{eq:ContinuumModel}:
\begin{equation}
\widetilde{H}_{\rm int} = g \int {\rm d}x \, \psi_+^\dagger \psi_-^\dagger \psi_- \psi_+ \, , \quad g = 2V_0 a^2/3 \, .
\end{equation}

\subsection{Binding Energy}

As a consistency check of our continuum model Eq.~\ref{eq:ContinuumModel}, we can determine the binding energy and compare it with the exact solution found in App.~\ref{app:TwoBodyProblem}. 
Using a $T$-matrix approach, the binding energy $\varepsilon_b$ of the continuum theory is solution of the implicit equation~\cite{levinsen2015strongly}
\begin{equation}
\frac{1}{g} = - \int \frac{{\rm d}^2 q}{(2\pi)^2} \frac{1}{\varepsilon_b + |q|^2/m} \, .
\end{equation}
Going to polar coordinates $q= q_r e^{i\theta}$ and introducing the momentum UV cutoff $\Lambda$, we find
\begin{equation}
\frac{2 \pi}{m |g|} = \int_0^{\Lambda} {\rm d}q_r \frac{q_r}{q_r^2 + m \varepsilon_b}  \, ,
\end{equation}
which can be integrated to obtain 
\begin{equation} \label{appeq:ContinuumBindingEnergy}
\varepsilon_b = \frac{\Lambda^2}{m \left[ e^{\frac{4\pi}{m|g|}} - 1 \right]} \, .
\end{equation}

In order to fix the cutoff $\Lambda$, we can use our solution of the problem in the limit $V \gg \Delta$ of Sec.~\ref{sapp:ResonatingTriangle}, $\varepsilon_b = 3 t^2 / \Delta$. In that limit, Eq.~\ref{appeq:ContinuumBindingEnergy} reduces to $\varepsilon_b \simeq (3\Lambda t)^2/(2\pi \Delta)$. Equating the two limits yields:
\begin{equation}
\Lambda = \sqrt{\frac{2\pi}{3a^2}} \simeq \frac{1.44720}{a} \, .
\end{equation}
In Eq.~\ref{eq:ContBinding}, we have introduce the corresponding energy cutoff $\varepsilon_{\rm uv} = \Lambda^2 / m = \pi t_f = \pi  W /9$.

\section{Longer range interactions} \label{app:LongerRangeInteraction}

In this appendix, we show evidence that longer range interactions neither destroy the effective pairing between doped charges nor the superconducting phase described in the main text. We include next-nearest neighbor interactions $\mathcal{H}_2 = V_2 \sum_{\langle\langle r, r' \rangle \rangle} n_r n_{r'}$ in order to describe the tail of the fermion-fermion interaction.

The range of next-nearest interaction strength relevant for our work is $\Delta \sim V > V_2$. In that regime, the Schrieffer-Wolff transformation of App.~\ref{app:SchriefferWolff} almost identically applies, up to two corrections. To first order, $\mathcal{H}_2$ introduces a direct interaction between electron in the $f$-band. To second order in $t$, it also changes the denominators of Eq.~\ref{eq:CoefficientSchriefferModel}, which are determined by the new dipole, polaron and trimer energies: 
\begin{equation} \begin{split}
E_D & = \Delta + 2V -6V_2 \, , \\ E_P & = E_f + \Delta + V - 5 V_2 \, , \\ E_T & = 2 E_f + \Delta - 3 V_2 \, .
\end{split} \end{equation}
For weak perturbations $V_2 \ll \Delta$, the second corrections can be safely discarded, and we obtain the same effective model as Eq.~\ref{eq:BelectronEffModel} with a two-body interaction strength 
\begin{equation}
V_f' = V_2 + V_f \, .
\end{equation}
At small doping, we can go to the continuum and the system is described by Eq.~\ref{eq:ContinuumModel} with a modified coupling constant $g' = 3a^2 (V_f' - 2\lambda)$ (compare with Eq.~\ref{couplingconstant}). When the additional interaction term is smaller than the largest pair binding energy $\varepsilon_b$ of the original problem, \textit{i.e.} $V_2 < 3 t^2/\Delta$, we find a that $g'$ is negative when $V/\Delta \geq x_+ $, where
\begin{equation}
x_+ = \frac{3 v_2 + \sqrt{v_2(v_2+24)}}{4(3-v_2)} \, , \quad v_2 = \frac{V_2 \Delta}{t^2} \, .
\end{equation}
In other words, at small $\delta$, longer range interactions in the range $V_2 < 3t^2/\Delta$ only shifts the range of $V/\Delta$ where superconductivity is observed. For larger $V_2$, the bound pairs breaks to avoid large next-nearest neighbor repulsion and the binding energy is always negative.

\begin{figure}
\centering
\includegraphics[width=\columnwidth]{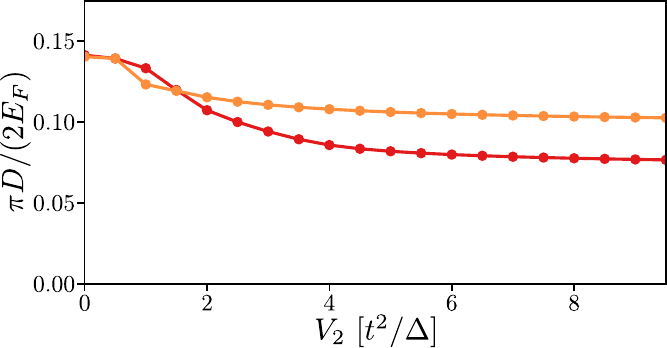}
\caption{\textbf{Charge stiffness as a function of $V_2$:}
for 4 (orange) and 6 (red) particles in a $7 \times 7$ lattice with $V = 0.6 \Delta$. 
}
\label{figapp:StableagainstV2}
\end{figure}

Even in that regime, the superconducting phase can reappear at larger doping concentrations. Indeed, if they are closely packed, the fermions of a pair do not decrease their energy by splitting apart because this send them closer to all the other surrounding fermions in the system. 
We evidence the stability of the superconducting phase against $V_2$ at large doping with ED with 4 and 6 particles in a $7 \times 7$ lattice, thus at doping $\delta = 4/(7\times7) \simeq 8\%$ and $6/(7\times 7) \simeq 12 \%$, with $V = 0.6 \Delta$. The charge stiffness extracted as a function of $V_2$ is shown in Fig.~\ref{figapp:StableagainstV2}. Despite a slight decrease, the stiffness remains positive up to $V_2 = 10 t^2/\Delta$. This gives support for the stability of the superconducting phase against longer range interactions at doping $\delta \sim 0.1 - 0.15$.

Finally, the regime $\Delta \sim V_2 \gg t$ has been extensively described in the in Ref.~\cite{slagle2020charge}. In that case, the trimer energy $E_T= 2 E_f + \Delta - 3 V_2$ can be resonant with the $f$-band. The presence of these preformed pairs in the Fermi sea of doped charges leads to superconductivity and pair-density waves. While the origin of pairs is different in that regime, strong $V_2$ interactions can still lead to superconductivity.

\section{Pairing and superconductivity in the full-fledged model} \label{app:EDresults}

\begin{figure}
\centering
\includegraphics[width=\columnwidth]{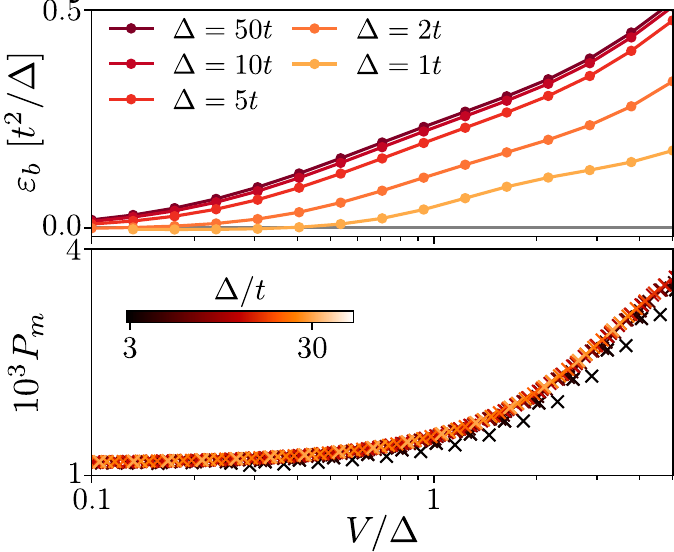}
\caption{\textbf{Pair binding energy and pair correlation function:}
shown for a finite honeycomb lattice with 30 sites. They highlight the presence of two-body bound state and of a superconducting phase in the original model.
}
\label{figapp:EDOriginalModel}
\end{figure}

We present ED results on the original model Eq.~\ref{eq:OriginalModel} in Fig.~\ref{figapp:EDOriginalModel}, which give further evidence for the presence of two-body bound pairs at finite doping above the $n=1$ insulating state, and for the emergence of superconductivity in our model. The former is probed by the pair binding energy $\varepsilon_b$ that we find, as in the main text, positive in the limit $t \ll \Delta$ for the entire range of $V/\Delta$ considered. For $\Delta \sim t$, we also observe a region with positive pair binding energy for sufficiently large $V/\Delta$. 

To investigate the superconducting correlations in the ground state of the doped $n=1$ insulator, we calculate the $f$-wave pair correlation function
\begin{equation}
P(r \in B) = \frac{1}{N_s} \sum_{x \in B} \left\langle \Delta_f^\dagger (r+x) \Delta_f (x) \right\rangle \, , 
\end{equation}
with $\Delta_f (x) = \sum_{j=1}^3 c_{x+\delta_j} c_x$. A superconducting state with $f$-wave symmetry is expected to exhibit a power-law decay of $P(r)$ with distance, and hence to have sizable correlations at large distances. For the largest accessible finite size clusters $N_s=5\times 3$, the maximum distance is achieved for $r_m = (2,2)$. We denote the pair correlation function at this distance as $P_m = P(r_m)$. In Fig.~\ref{figapp:EDOriginalModel}, we show $P_m$ as a function of the parameters $V/\Delta$ for several $\Delta/t$. We observe large correlations in the perturbative limit $\Delta>5t$ when $V/\Delta > 0.25$, where we indeed expect a strong superconducting order (see, for instance, Fig.~\ref{fig:CriticalTemperature}). This direct manifestation of superconductivity gives strong evidence for the reliability of the predictions made from the effective model Eq.~\ref{eq:BelectronEffModel} and its mean-field solution detailed below in App.~\ref{app:Meanfieldsuperconductivity}.

\section{Mean-field treatment} \label{app:Meanfieldsuperconductivity}

\begin{figure}
\centering
\includegraphics[width=\linewidth]{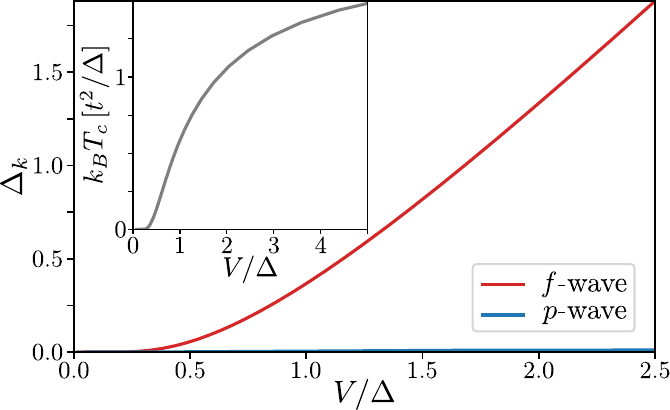}
\caption{ \textbf{Numerical solution of the gap equation:}
Numerical solution $\Delta_k$ of the self-consistent gap equation for infinitesimal doping, decomposed into partial waves of $p$ and $f$ symmetry. The inset shows the self-consistent estimate of the critical temperature for $\delta=0.05$, which almost perfectly match the exact formula Eq.~\ref{eq:BCSCriticalTemp}.
}
\label{figapp:BcsSolutionGapEquation}
\end{figure}

\subsection{Dilute limit}

We now carry out a self-consistent mean field treatment of the low-energy model Eq.~\ref{eq:BelectronEffModel} in the dilute limit, where three-body interaction terms are negligible. Going to Fourier space, the Hamiltonian becomes
\begin{equation} 
\mathcal{H}' = \sum_k \varepsilon_k f_k^\dagger f_{k} + \frac{1}{N_s} \sum_{k,q,q'} V_{k,q} f_{q-k}^\dagger f_{q'+k}^\dagger f_{q'} f_{q} \, ,
\end{equation}
with $V_{k,q}$ defined in Eq.~\ref{appeq:InteractionMomentumSpace}. Then we perform the mean-field substitution $f_{q'} f_{q}  \simeq \delta_{q+q'} \langle f_{-q} f_{q} \rangle$. This replacement leads to the following quadratic mean-field Hamiltonian
\begin{equation}  \begin{split}
\mathcal{H}_{\rm MF} & = \frac{1}{2} \sum_k \begin{pmatrix} f_k^\dagger & f_{-k} \end{pmatrix} \begin{pmatrix} \xi_k & \Delta_k \\ \Delta_k^* & - \xi_k \end{pmatrix} \begin{pmatrix} f_k \\ f_{-k}^\dagger \end{pmatrix} \, , \\ \Delta_k & = \frac{2}{N_s} \sum_q \Re(V_{q-k, q}) \langle f_{-q} f_q \rangle \, ,
\end{split} \end{equation}
where we have introduced a chemical potential $\mu$ and the short notation $\xi_k = \varepsilon_k - \mu$. We have also used the relation $V_{-k,-q} = V_{k,q}^*$. The quadratic Hamiltonian $\mathcal{H}_{\rm MF}$ is diagonalized as
\begin{equation}
\mathcal{H}_{\rm MF} = \sum_k E_k \gamma_k^\dagger \gamma_k \, , \quad E_k = \sqrt{\xi_k^2 + |\Delta_k|^2} \, ,
\end{equation}
through the Bogoliubov transformation
\begin{align}
\begin{pmatrix} f_k \\ f_{-k}^\dagger \end{pmatrix} & = \begin{pmatrix}
	u_k & v_k \\ -v_k^* & u_k^*
\end{pmatrix} \begin{pmatrix} \gamma_k \\ \gamma_{-k}^\dagger \end{pmatrix} \, , \\ 
u_k & = \sqrt{\frac{1}{2} \left( 1+\frac{\xi_k}{E_k}\right)} \, , \,\, v_k = - \frac{\Delta_k}{|\Delta_k|} \sqrt{\frac{1}{2} \left( 1-\frac{\xi_k}{E_k}\right)} \, . \notag
\end{align}
The ground state of the system is annihilated by all quasi-particle operators $\gamma_k$ and takes the form $ \ket{\Psi_{\rm BCS}} = \prod_k \gamma_k \ket{n=1} $.
The order parameter $\Delta_k$ may now be computed self-consistently through the calculation of the correlators $\braOket{\Psi_{\rm BCS}}{f_{-q} f_q }{\Psi_{\rm BCS}}$. This gives the so-called gap equation
\begin{equation}  \begin{split}
\Delta_k & = \frac{1}{N_s} \sum_q  \Re( V_{q-k,q} - V_{q+k,q} ) u_q^* v_q \\
& = \frac{1}{2 N_s} \sum_q  \Re( V_{q+k,q} - V_{q-k,q} ) \frac{\Delta_q}{E_q} \, .
\end{split} \end{equation}
Using the explicit form of $V_{k,q}$, this gap equation can be rewritten as 
\begin{equation}
\Delta_k = \sum_{j=1}^3 u_j \sin(k\cdot a_j) \, , \,\, u_j = \frac{1}{N_s} \sum_q \phi_j(q) \frac{\Delta_q}{E_q} \, ,
\end{equation}
with
\begin{equation}
\phi_j(q) = -V_f \sin (q a_j) + \lambda [\sin (q a_{j+1})+\sin (q a_{j-1})] \, .
\end{equation}
Similarly, the mean particle number fixes the chemical potential $\mu$ through the relation
\begin{equation}
\delta = \frac{1}{2N_s} \sum_q \left( 1 - \frac{\xi_q}{E_q} \right) \, .
\end{equation}

We solve the self-consistent gap equation numerically for infinitesimal doping and decompose the order parameter $\Delta_k$ into partial $p$ and $f$-wave. The results are shown in Fig.~\ref{figapp:BcsSolutionGapEquation}. We observe that the self-consistent always has a solution, with a substantial gap for $\Delta/V > 0.25$. This provides another test of the continuum results of Fig.~\ref{fig:CriticalTemperature}, where the critical temperature drastically rises near $\Delta/V = 0.25$. Furthermore, our mean-field solution exhibits a strong $f$-wave symmetry, which corroborates our exact two-fermion calculation (Fig.~\ref{fig:PairBindingBandwidth}).

\subsection{Critical temperature and and three-body interactions}

Including finite temperature effects to the previous analysis provides a way to determine the critical temperature $T_c$ at which the pair are all broken by thermal fluctuation and $\Delta_k=0$~\cite{schrieffer2018theory}. This definition agrees with the superconducting temperature in the BCS limit of our model. Note that it does not capture the physics of the BKT transition, explaining why we obtain $k_B T_c \geq E_F/8$ in Fig.~\ref{fig:CriticalTemperature}. Focusing on an order parameter with pure $f$-wave symmetry, the gap and number equations become
\begin{subequations} \label{eqapp:TcDeterminationBCS}
\begin{align} 
1 & = \frac{1}{N_s} \sum_q \phi_j(q) \frac{\sum_p \sin(q \cdot a_p)}{|\xi_q|} \tanh \left(\frac{|\xi_q|}{2k_B T_c}\right) \\
\delta & = \frac{1}{2 N_s} \sum_q \left[ 1 - \frac{\xi_q}{|\xi_q|} \tanh \left(\frac{|\xi_q|}{2k_B T_c}\right) \right] \, . 
\end{align}
\end{subequations}
We solve this implicit definition of $T_c$ numerically. The results, presented in the inset of Fig.~\ref{figapp:BcsSolutionGapEquation} for $\delta = 0.05$, show a sharp increase of $T_c$ near $V = \Delta/4$. Excellent agreement is found with the continuum results of Fig.~\ref{fig:CriticalTemperature} in the small doping limit. For $\delta \geq 0.1$, the non-parabolicity of the dispersion relation leads to corrections to Eq.~\ref{eq:BCSCriticalTemp}.

The three-body interaction terms of Eq.~\ref{eq:BelectronEffModel} can be partially reincorporated in Eq.~\ref{eqapp:TcDeterminationBCS} near $T_c$ via the substitution 
\begin{equation*}
n_i n_j n_k \simeq \delta (n_i n_j + n_j n_k + n_i n_k) - \beta ( f_i^\dagger n_j f_k + P_{ijk} ) \, ,
\end{equation*}
where we have replace one quadratic operator by its expectation in the normal state $\langle n \rangle = \delta$ and $\langle f_k^\dagger f_i \rangle = \beta  = N_s^{-1} \sum_k \cos(k\cdot a_1) f_{\rm FD} (\xi_k)$ with $f_{\rm FD}$ the Fermi-Dirac distribution. This simply renormalizes the effective interaction $V_f$ and correlated hopping strength
\begin{equation}
V_f' = V_f + \delta U_3 \, , \quad \lambda' = \lambda - U_3 \beta \, ,
\end{equation}
which leads to a similar replacement of $\phi_j$ by 
\begin{equation}
\phi_j' (q) = - V_f' \sin (q a_j) + \lambda' [\sin (q a_{j+1})+\sin (q a_{j-1})] 
\end{equation}
in Eq.~\ref{eqapp:TcDeterminationBCS}, as declared in the main text. The results shown in Fig.~\ref{fig:CriticalTemperature} are obtained with this substitution.

For fillings $\delta > 1/4$, we must rely on Eq.~\ref{eqapp:TcDeterminationBCS}, which correctly describes the merging of the $\pm K$ pockets, while the two-flavor continuum model Eq.~\ref{eq:ContinuumModel} assume them well-separated. At these doping concentrations, our calculations show that a superconducting state only appears for $V \geq 0.3 \Delta$, in sharp contrast with the dilute regime where superconductivity extends to $V\ll \Delta$ (see Fig.~\ref{fig:ChargeStiffnessNumerics}). Another stark difference with the dilute limit is the presence of nodes of the superconducting gap located at the Fermi surface, which now encloses the $\Gamma$ point.

\begin{figure}
\centering
\includegraphics[width=\linewidth]{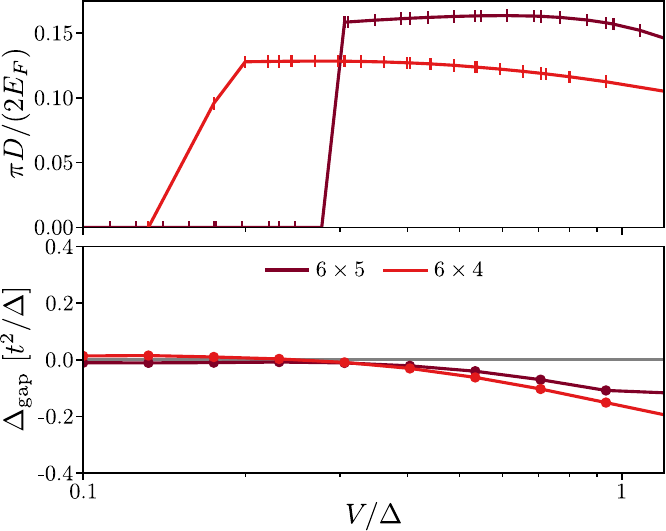}
\caption{ \textbf{Phase stiffness and superconducting gap:} obtained by ED of Eq.~\ref{eq:BCSCriticalTemp} for 8 particles in $6\times5$ (brown) and $6\times4$ (red) finite lattices.
}
\label{figapp:NodeSuperconductorED}
\end{figure}

\section{ED evidence for nodal superconductor} \label{app:NodeSCFromED}

We present more evidence of this nodal superconductor from ED of the effective model for $\delta > 1/4$. We focus on lattices of size $6\times4$ and $6\times5$ with 8 particles and more, and extract the charge stiffness and the superconducting gap as in the main text. Our results are presented in Fig.~\ref{figapp:NodeSuperconductorED}. There, we observe that the system is not superconducting for $V \ll \Delta$, in agreement with the mean-field calculations of App.~\ref{app:Meanfieldsuperconductivity}. The transition to $D>0$ occurs around $V = 0.2 \Delta$, consistent with the mean-field value up to finite-size jitters. These observations comfort the presence of a superconducting phase at doping $\delta > 1/4$ and moderate $V/\Delta$. 

In the entire superconducting phase evidenced above, the superconducting gap $\Delta_{\rm gap}$ is almost zero and decrease in magnitude with larger system size. This hints towards nodes of the superconducting gap at the Fermi energy, as expected from the Fermi surface topology and $f$-wave symmetry of the order parameter (see text).

\section{Bosonic effective interactions and phase separation} \label{app:BosonicEffectiveInt}

\begin{figure}
\centering
\includegraphics[width=\linewidth]{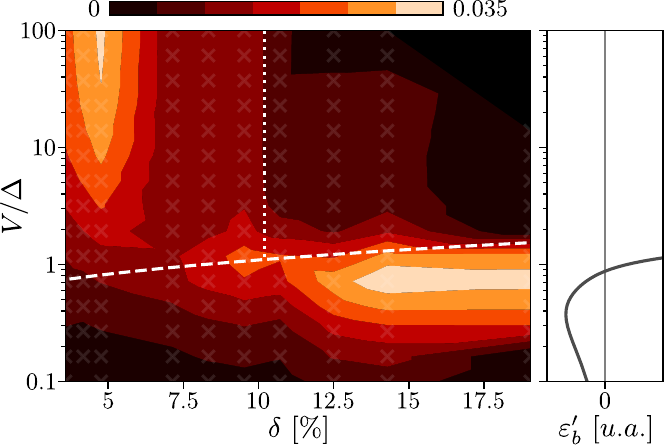}
\caption{ \textbf{
Phase stiffness in a wider region of doping and ratio $\bm{V/\Delta}$. The white dashed line distinguishes between the BCS and BEC limits:} The right panel shows the second binding energy. Its positivity indicates effective attractions between bosons deep in the BEC phase. 
}
\label{figapp:PhaseDiag}
\end{figure}

In this appendix, we present our numerical results obtained outside of the superconducting region that we focus on in the main text. We have evaluated the charges stiffness of the ground state in a wide region of doping concentration and ratio $V/\Delta$. Our results are presented in Fig.~\ref{figapp:PhaseDiag} for lattices with $L_1 \geq 7$ and $L_2 \geq 6$, which mitigate finite size effects. A white dashed line indicates where $\varepsilon_b = E_F$, and serves as an indicator to distinguish between the BCS and BEC limits predicted by the continuum model Eq.~\ref{eq:ContinuumModel}. As shown in the main text, the stiffness in the BCS region ($V<\Delta$) slowly increases but remains close to $E_F / 8$, especially at small doping concentrations. In the BEC limit, we observe two different behaviors depending on the value of $\delta$. First, the results at $\delta < 10\%$ cannot reliably describe the thermodynamic behavior of the system as they were obtain for less than four particles. In that case, the small BEC is either a two-body or four-body bound state with a large effective mass. For larger doping, the lattice effects observed in the main text rule the physics and the charge stiffness drops down with increasing $V/\Delta$. We observe that this diminution comes with a reduction of the may-body spectrum gap, until the ground state becomes highly degenerate for the $V > 70 \Delta$. 

\begin{figure}
\centering
\includegraphics[width=\linewidth]{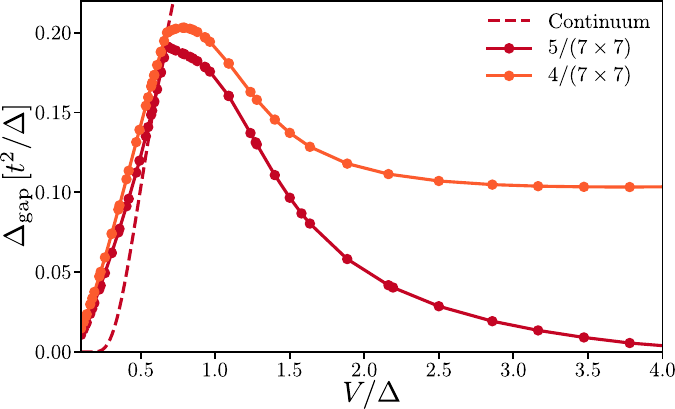}
\caption{ \textbf{
Superconducting gap in a wider range of $\bm{V/\Delta}$:} It reduces to almost zero at $V=4\Delta$ for large doping concentrations. 
}
\label{figapp:SCgapLargeScale}
\end{figure}

To understand these features, we investigate the scattering properties of the local bosons in the $V\gg \Delta$ limit. To distinguish between attractive and repulsive interactions of pairs, we evaluate the second binding energy $\varepsilon_b' = 2 E(2) - E(4) - E(0)$, which plays the same role as $\varepsilon_b$ for the tightly bound pairs. The results presented to the right of Fig.~\ref{figapp:PhaseDiag}, show a positive $\varepsilon_b'$ for $V > \Delta$. This signals attractive interaction, which leads to the collapse of the quasi-BEC and to phase-separation. The collapse of the BEC in favor of phase separation explains the large degeneracy observed for $V \gg \Delta$. The BEC collapse at large doping concentrations for $V>\Delta$ is also evidenced by a decay of the superconducting gap for $V>\Delta$, as shown in Fig.~\ref{figapp:SCgapLargeScale}. Note that the points in the range $V<1.2\Delta$ reproduce those of Fig.~\ref{fig:ChargeStiffnessNumerics} in the main text.

\newpage

\end{document}